\documentclass[journal]{IEEEtran}
\usepackage{cite}
\usepackage[cmex10]{amsmath} 
\usepackage{amssymb}
\usepackage{amsfonts}
\usepackage{amsthm} 
\usepackage{algorithmic} 
\usepackage{algorithm}
\usepackage{graphicx}
\usepackage{textcomp}
\usepackage{xcolor} 
\usepackage{bm}
\usepackage{booktabs}
\usepackage{multirow} 

\begin{document}

\title{Real Interference Alignment for Active IRS-Aided Systems: A Rate-Profile Learning-Based Approach}

\author{Junda Liao, Quanzhong Li, and Qi Zhang, \emph{Member}, \emph{IEEE}

\thanks{\emph{(Corresponding author: Qi Zhang.)}

Junda Liao and Qi Zhang are with the School of Electronics and Information Technology, Sun Yat-sen University, Guangzhou 510006, China (e-mail: liaojd3@mail2.sysu.edu.cn; zhqi26@mail.sysu.edu.cn). Quanzhong Li is with the School of Computer Science and Engineering, Sun Yat-sen University, Guangzhou 510006, China (e-mail: liquanzh@mail.sysu.edu.cn).}

}

\markboth{}{Liao \textit{et al.}: Real IA for Active IRS-Aided Systems}%

\maketitle

\begin{abstract}
With additional spatial degrees of freedom provided by the active intelligent reflecting surface (IRS), interference alignment (IA) can be achieved at low cost. In this letter, we propose a real IA scheme for an active IRS-aided system. The proposed scheme only requires the IRS to know the instantaneous channel coefficients under the assumption of blocked direct links. To maximize the achievable sum rate subject to individual minimum rate requirements and transmission power constraints, we propose a rate-profile learning-based algorithm. The algorithm uses offline-trained achievable rate profiles to decouple the original problem into multiple feasibility subproblems, which are then solved by generalized eigenvalue decomposition. Simulation results demonstrate that our proposed algorithm outperforms the conventional weighted minimum mean square error algorithm, while requiring significantly less program execution time.
\end{abstract}

\begin{IEEEkeywords}
Active intelligent reflecting surface (IRS), generalized eigenvalue decomposition, interference alignment (IA), weighted minimum mean square error (WMMSE).
\end{IEEEkeywords}

\section{Introduction}

Interference alignment (IA) confines interference to a reduced-dimensional subspace while preserving interference-free dimensions for desired signals \cite{Cadambe08}. For interference channels, real-domain signaling provides additional structure for IA \cite{Motahari14}, while asymmetric complex signaling exploits the real representation of complex channels \cite{Cadambe10}. An intelligent reflecting surface (IRS) provides additional spatial degrees of freedom and is therefore attractive for facilitating IA \cite{Wu20,Long21,JYao26}.

Unlike a passive IRS, an active IRS amplifies reflected signals to alleviate cascaded-link attenuation, at the cost of amplified noise and power constraints \cite{Long21,JYao26,Zhang23}. The passive IRS-assisted IA was studied in \cite{MFu21}. For active IRS interference channels, Long \emph{et al.} proposed a general alternating optimization framework to obtain the achievable rate region \cite{Long24}. A joint IA precoding and active IRS beamforming scheme was proposed in \cite{Liu26} to achieve the concurrent transmission of multiple interference-free data streams. 

Achieving IA in both passive and active IRS-aided systems \cite{MFu21,Liu26} requires the transmitters, the IRS, and the receivers to know the instantaneous channel coefficients, which is a costly requirement. In this letter, we propose a real IA scheme for an active IRS-aided system. Assuming the direct links between each single-antenna transmitter-receiver pair are blocked by obstacles, the proposed scheme only requires the IRS to know the instantaneous channel coefficients. 

For the proposed system, we formulate an achievable sum rate maximization problem subject to both the minimum achievable rate requirements for each transmitter-receiver pair and the transmission power constraints. To solve the formulated problem, the weighted minimum mean square error (WMMSE) algorithm is commonly employed \cite{Shi11}. However, the WMMSE algorithm may converge to a local optimum, and its performance is highly sensitive to the initial point. In this letter, we propose a rate-profile learning-based algorithm. The algorithm uses the offline-trained achievable rate profiles to decouple the original problem into multiple feasibility check problems. Each problem is then solved via generalized eigenvalue decomposition.

\emph{Notations}: Boldface lowercase and uppercase letters denote vectors and matrices, respectively. $\mathbf{A}^T$, $\mathbf{A}^\ast$, and $\mathbf{A}^H$ denote the transpose, conjugate, and conjugate transpose of the matrix $\mathbf{A}$, respectively. The Hadamard product is denoted as $\odot$. $\mathbf{e}_n$ denotes the standard basis vector whose $n$th element is one and all other elements are zero. $\mathbb R$ and $\mathbb C$ denote the sets of real and complex numbers, respectively.

\section{System Model}

Consider an active IRS-aided multi-user interference channel, which consists of $K$ single-antenna transmitter-receiver pairs. The direct link between any transmitter and any receiver is blocked by obstacles. To maintain connectivity, the active IRS equipped with $N$ reflecting elements is deployed to establish reliable wireless communications by amplifying the incident signals from transmitters and reflecting them towards receivers.

In the proposed system, the $k$th transmitter sends a real-valued symbol $\nu_k$ with $\mathbb{E}[\nu_k^2] = P_{\nu}$ directly without applying any channel-dependent precoding to the $k$th receiver, $k\in\mathcal{K}=\{1, 2, \cdots, K\}$. Denote the baseband channel vector from the $k$th transmitter to the IRS as $\mathbf{g}_k=(g_{1,k},\cdots,g_{N,k})^T\in \mathbb{C}^{N \times 1}$ and that from the IRS to the $k$th receiver as $\mathbf{h}_{k}^H=(h_{1,k},\cdots,h_{N,k}) \in \mathbb{C}^{1 \times N}$. The received signal at the $j$th receiver, $j\in\mathcal{K}$, is written as
\begin{align}\label{q1}
y_{j}&=\mathbf{h}_{j}^H \mathbf{\Theta} \sum_{k=1}^K\mathbf{g}_k \nu_k+ \mathbf{h}_{j}^H \mathbf{\Theta}\mathbf{n} + w_{j}
\end{align}
where $\mathbf{n}\sim \mathcal{CN}(\mathbf{0}, \sigma_{\text{IRS}}^2 \mathbf{I}_N)$ denotes the input-referred dynamic active noise introduced by the IRS elements, $w_{j} \sim \mathcal{CN}(0, \sigma^2)$ denotes the thermal noise at the receiver, and
\begin{align}
\mathbf{\Theta}=\text{diag}(\bm{\theta})=\text{diag}(\theta_1, \theta_2, \cdots, \theta_N)^T
\end{align}
denotes the active beamforming matrix at the IRS.

Using the property 
\begin{equation}
\mathbf{a}_1^H \text{diag}(\mathbf{a}_2) \mathbf{a}_3 = \mathbf{a}_2^T \left(\mathbf{a}_1^* \odot \mathbf{a}_3\right),
\end{equation}
we rewrite \eqref{q1} as
\begin{align}
y_{j}&=\bm{\theta}^T \sum_{k=1}^K\mathbf{f}_{j,k}\nu_k+ \tilde{w}_{j}
\end{align}
where $\mathbf{f}_{j,k}=\mathbf{h}_{j}^* \odot \mathbf{g}_k$ denotes the cascaded channel and 
\begin{align}
\tilde{w}_{j}&=\mathbf{h}_{j}^H \mathbf{\Theta}\mathbf{n} + w_{j}
\end{align}
denotes the aggregate effective noise.

Assuming that $\mathbf{g}_k$ and $\mathbf{h}_{k}$, $k\in\mathcal{K}$, are known at the IRS, our objective in this letter is to design the beamforming vector $\bm{\theta}$ to achieve real IA \cite{Motahari14}. For the $k$th transmitter-receiver pair, $k\in\mathcal{K}$, the condition for interference-free reception in the real domain is
\begin{equation}\label{q8}
\text{Re}\left\{\bm{\theta}^T \mathbf{f}_{j,k} \right\} = 0,\ \forall\ j \neq k.
\end{equation}
By satisfying \eqref{q8}, all inter-user interference is aligned into the imaginary dimension, allowing the receiver to recover the desired real-valued symbol $\nu_k$ by taking the real part of $y_k$, free from interference.

Define 
\begin{align}
\mathbf{b}_{j,k}&=(\text{Re}(\mathbf{f}_{j,k})^T,-\text{Im}(\mathbf{f}_{j,k})^T)^T,\\
\mathbf{x}&=(\text{Re}(\bm{\theta})^T,\text{Im}(\bm{\theta})^T)^T,\\
\mathbf D_j&=\text{diag}(|\mathbf h_j|^2,|\mathbf h_j|^2).
\end{align}
The real cascaded coefficient is $\mathbf b_{j,k}^T\mathbf x$, and the noise variance for real part detection is $(\sigma_{\text{IRS}}^2\mathbf{x}^T\mathbf{D}_j\mathbf{x}+\sigma^2)/2$. From \eqref{q8}, we know that the interference-free reception condition is $\mathbf b_{j,k}^T\mathbf x=0$ for all $j\ne k$. Stacking $\mathbf b_{j,k}^T$ for all $j\ne k$ gives $\mathbf{B}$. Let $\mathbf Z\in\mathbb{R}^{2N\times d}$ be the nullspace of $\mathbf{B}$, where $d=2N-K(K-1)>0$. The IA constraint is imposed exactly by 
\begin{equation}
\mathbf{x}=\mathbf{Z}\mathbf{s}
\end{equation}
where $\mathbf{s}\in\mathbb{R}^{d\times1}$ denotes the real beamforming vector to be optimized. 

After IA, the signal-to-noise ratio (SNR) of the $k$th receiver, $k\in\mathcal{K}$, is 
\begin{equation}\label{q11}
\gamma_k=\frac{P_{\nu}\left(\mathbf{c}_k^T\mathbf{s}\right)^2}{\mathbf{s}^T\mathbf{M}_k\mathbf{s}+\sigma^2/2}
\end{equation}  
where $\mathbf{c}_k=\mathbf{Z}^T\mathbf{b}_{k,k}$ and $\mathbf{M}_k=\sigma_{\text{IRS}}^2\mathbf{Z}^T\mathbf{D}_k\mathbf{Z}/2$. The corresponding achievable rate is
\begin{equation}\label{q12}
R_k=\frac{1}{2}\log_2(1+\gamma_k)
\end{equation}  
where the factor $\frac{1}{2}$ is included because only the real part of the received signal is used for signal detection.

The transmission power of the $n$th reflecting element of the IRS, $n\in\mathcal{N}=\{1, 2, \cdots, N\}$, is $\mathbf{s}^T\mathbf{Q}_n\mathbf{s}$, where
\begin{align}\label{q13}
\mathbf{Q}_n&=q_n\mathbf{Z}^T\left(\mathbf{e}_n\mathbf{e}_n^T+\mathbf{e}_{N+n}\mathbf{e}_{N+n}^T\right)\mathbf Z,\\
\label{q14}q_n&=\sigma_{\text{IRS}}^2+P_{\nu}\sum_{k=1}^K|g_{n,k}|^2.
\end{align}
The total transmission power of the IRS is $\mathbf{s}^T\mathbf{P}\mathbf{s}$, where
\begin{align}\label{q15}
\mathbf{P}=\mathbf{Z}^T\sum_{n=1}^N q_n\left(\mathbf{e}_n\mathbf{e}_n^T+\mathbf{e}_{N+n}\mathbf{e}_{N+n}^T\right)\mathbf{Z}.
\end{align}
Thus, the achievable sum rate maximization problem of the multi-user interference channel is formulated as 
\begin{subequations}\label{q20} 
\begin{align}
\label{q20a}\max_{\mathbf{s}\in\mathbb{R}^{d\times1}}\ & R_{\text{sum}}=\frac{1}{2}\sum_{k=1}^K\log_2(1+\gamma_k)\\
\label{q20b}\text{s.t.}\ \ &\frac{1}{2}\log_2(1+\gamma_k)\geq R_{\min},\ \forall\ k\in\mathcal{K},\\
\label{q20c}&\mathbf{s}^T\mathbf{P}\mathbf{s}\leq P_t,\ \mathbf{s}^T\mathbf{Q}_n\mathbf{s}\leq P_e,\ \forall\ n\in\mathcal{N}
\end{align}
\end{subequations}
where $R_{\min}$ denotes the minimum achievable rate requirement for each transmitter-receiver pair, $P_t$ and $P_e$ denote the total and per-element transmission power constraints at the IRS, respectively. 

\section{Rate-Profile Learning-Based Algorithm}

Problem \eqref{q20} can be solved by the WMMSE algorithm. However, the WMMSE algorithm involves triple alternating optimization, resulting in a large number of iterations for convergence. In this letter, we propose a rate-profile learning-based algorithm to solve problem \eqref{q20}.

\subsection{Offline Training}

Let $M$ denote the number of training channels. For the $m$th training channel, $m\in\mathcal{M}=\{1, 2, \cdots, M\}$, 
we employ the WMMSE algorithm to obtain the solution to problem \eqref{q20}, denoted as $\mathbf{s}^{(m)}$. 
In this letter, we define the desired-link strengths which are projected onto the nullspace of $\mathbf{B}$ as
\begin{equation}
\rho_k=\mathbf{c}_k^T\left(\mathbf{M}_k+\sigma^2/2\mathbf{I}\right)^{-1}\mathbf{c}_k.
\end{equation}
For the $m$th training channel, we sort $\rho_k$ such that 
\begin{equation}\label{bq2}
\rho_{\pi_m(1)}^{(m)}\le\cdots\le\rho_{\pi_m(K)}^{(m)}
\end{equation}
where $\bm{\pi}_m=(\pi_m(1),\cdots,\pi_m(K))^T$ denotes a permutation of the elements in $\mathcal{K}$ to satisfy \eqref{bq2}.

In problem \eqref{q20}, the gradient of the objective function is
\begin{equation}\label{bq3}
\nabla_{\mathbf{s}} R_{\text{sum}}^{(m)}=\sum_{k=1}^K\nabla_{\mathbf{s}} R_{\pi_m(k)}^{(m)}.
\end{equation}
Substituting \eqref{q11} and \eqref{q12} into \eqref{bq3}, we obtain
\begin{align}\label{bq4}
\nabla_{\mathbf{s}} R_{\pi_m(k)}^{(m)}=&\omega_{\pi_m(k)}^{(m)}P_{\nu}\mathbf{c}_{\pi_m(k)}\mathbf{c}_{\pi_m(k)}^T\mathbf{s}^{(m)}\nonumber\\
&-\omega_{\pi_m(k)}^{(m)}\gamma_{\pi_m(k)}^{(m)}\mathbf{M}_{\pi_m(k)}\mathbf{s}^{(m)}
\end{align}
where $\gamma_{\pi_m(k)}^{(m)}$ denotes the SNR of the $\pi_m(k)$th receiver for the $m$th training channel and
\begin{align}
\omega_{\pi_m(k)}^{(m)}&=\frac{1}{\zeta_{\pi_m(k)}^{(m)}\left(1+\gamma_{\pi_m(k)}^{(m)}\right)\ln2},\\
\zeta_{\pi_m(k)}^{(m)}&={\mathbf{s}^{(m)}}^T\mathbf{M}_{\pi_m(k)}\mathbf{s}^{(m)}+\sigma^2/2.
\end{align}
If the WMMSE algorithm cannot find a feasible solution to problem \eqref{q20}, this means that at least one of the minimum achievable rate requirements in \eqref{q20b} cannot be satisfied under total and per-element transmission power constraints. Under this condition, we set $\mathbf{s}^{(m)}=\mathbf{1}$ and $\gamma_{\pi_m(k)}^{(m)}=2^{2R_{\min}}-1$ to obtain $\omega_{\pi_m(k)}^{(m)}$.

With the offline-trained SNR, the achievable sum rate maximization problem \eqref{q20} is reduced to $M$ feasibility check problems, each expressed as follows 
\begin{subequations}\label{bq5} 
\begin{align}
\label{bq5a}\text{Find}_{\mathbf{s}\in\mathbb{R}^{d\times1}}\ & \frac{P_{\nu}\left(\mathbf{c}_k^T\mathbf{s}\right)^2}{\mathbf{s}^T\mathbf{M}_k\mathbf{s}+\sigma^2/2}\geq \gamma_k,\ \forall\ k\in\mathcal{K},\\
\label{bq5b}&\mathbf{s}^T\mathbf{P}\mathbf{s}\leq P_t,\ \mathbf{s}^T\mathbf{Q}_n\mathbf{s}\leq P_e,\ \forall\ n\in\mathcal{N}
\end{align}
\end{subequations}
where $\rho_{1}\le\cdots\le\rho_{K}$ is assumed. After optimization, we select the solution $\mathbf{s}$ that achieves the maximum sum rate among those with feasible solutions.

\subsection{Feasibility Check Solution}

In this subsection, we provide the solution to problems \eqref{bq5}. To derive the solution, we need the following proposition.

\emph{Proposition 1}: For the solution to problem \eqref{bq5}, at most a single constraint from the set of total and per-element transmission power constraints in \eqref{bq5b} is active, i.e.,
\begin{align}
\mathbf{s}^T\mathbf{P}\mathbf{s}=P_t \text{ or } \mathbf{s}^T\mathbf{Q}_n\mathbf{s}= P_e
\end{align}
for a specific $n$ among $\mathcal{N}$.

\emph{Proof}: See Appendix A. $\hfill\blacksquare$

Using Proposition 1, we assume that only the total power constraint is active and rewrite problem \eqref{bq5} as 
\begin{align}\label{bq10}
\text{Find}_{\mathbf{s}\in\mathbb{R}^{d\times1}}\ & P_{\nu}\left(\mathbf{c}_k^T\mathbf{s}\right)^2-\gamma_k\left(\mathbf{s}^T\mathbf{M}_k\mathbf{s}+\frac{\sigma^2}{2P_t}\mathbf{s}^T\mathbf{P}\mathbf{s}\right)
\geq 0, \nonumber\\
&\forall\ k\in\mathcal{K},\ \mathbf{s}^T\mathbf{P}\mathbf{s}=P_t.
\end{align}
By omitting the power constraint, the Lagrangian associated with problem \eqref{bq10} is given by
\begin{align}
\mathcal{L}=&\sum_{k=1}^K\omega_k\left(P_{\nu}\left(\mathbf{c}_k^T\mathbf{s}\right)^2-\gamma_k\mathbf{s}^T\left(\mathbf{M}_k+\frac{\sigma^2}{2P_t}\mathbf{P}\right)\mathbf{s}\right)
\end{align}
where $\omega_k>0$, $k\in\mathcal{K}$, denotes the Lagrange dual variable for constraints \eqref{bq5a}. Taking the first-order partial derivative of the Lagrangian with respect to $\mathbf{s}$ yields
\begin{align}\label{bq12}
\frac{\partial\mathcal{L}}{\partial\mathbf{s} }=&2\mathbf{C}\mathbf{s}-2\mathbf{D}\mathbf{s}
\end{align}
where $\mathbf{C}=\sum_{k=1}^K\omega_kP_{\nu}\mathbf{c}_k\mathbf{c}_k^T$ and
\begin{align}
\mathbf{D}=\sum_{k=1}^K\omega_k\gamma_k\left(\mathbf{M}_k+\frac{\sigma^2}{2P_t}\mathbf{P}\right).
\end{align}
The Karush–Kuhn–Tucker (KKT) conditions provide us with
\begin{equation}\label{bq15}
\mathbf{C}\mathbf{s}=\mathbf{D}\mathbf{s}.
\end{equation}
Multiplying both sides of \eqref{bq15} by $\mathbf{s}^T$, we obtain
\begin{equation}
\frac{\mathbf{s}^T\mathbf{C}\mathbf{s}}{\mathbf{s}^T\mathbf{D}\mathbf{s}}=1.
\end{equation}
Thus, the closed-form solution is the generalized eigenvector of the matrix pair $(\mathbf{C}, \mathbf{D})$. For the offline-trained SNR of the $m$th training channel, we propose to replace $\omega_{k}$ in the expression of $\mathbf{C}$ and $\mathbf{D}$ with $\omega_{\pi_m(k)}^{(m)}$ obtained in the previous subsection. This is because the expression $\frac{\partial\mathcal{L}}{\partial\mathbf{s}}$ in \eqref{bq12} is proportional to $\nabla_{\mathbf{s}} R_{\text{sum}}^{(m)}$ in \eqref{bq4}. 

Because $\mathbf{C}\succeq\mathbf{0}$ and $\mathbf{D}\succ\mathbf{0}$ have the size of $d\times d$, the number of generalized eigenvectors is $d$. Each obtained vector can be proportionally scaled to satisfy the total transmission power constraint. The result is denoted as $\mathbf{s}^o$. If the obtained $\mathbf{s}^o$ satisfies all the per-element transmission power constraints, it is a possible solution to problem \eqref{bq5}. Otherwise, because of Proposition 1, we proportionally scale $\mathbf{s}^o$ to be $\tau\mathbf{s}^o$, where
\begin{equation}
\tau=\min_{n\in\mathcal{N}} \left(\frac{P_e}{{\mathbf{s}^o}^T\mathbf{Q}_n\mathbf{s}^o}\right)^{\frac{1}{2}}.
\end{equation}
For each candidate vector, $\mathbf{s}^o$ or $\tau\mathbf{s}^o$, we test the feasibility against the constraints in \eqref{bq5a}. The achievable sum rate is computed only if all constraints are satisfied; otherwise, it is zero.
After solving $M$ feasibility check problems, we select the generalized eigenvector which corresponds to the largest achievable sum rate as the solution to problem \eqref{q20}. 

The whole rate-profile learning-based algorithm is summarized in Algorithm 1.

\begin{algorithm}
\caption{The Rate-Profile Learning-Based Algorithm}
\begin{algorithmic}[1]
\STATE For $M$ offline training channels, obtain $\{\gamma_{\pi_m(k)}^{(m)},\omega_{\pi_m(k)}^{(m)}\}$;
\STATE Given a multi-user interference channel, sort transmitter-receiver pairs such that $\rho_{1}\le\cdots\le\rho_{K}$;
\STATE \textbf{For} $m=1,\cdots,M$
\STATE \quad\quad Obtain $d$ generalized eigenvectors of $(\mathbf{C}, \mathbf{D})$;
\STATE \quad\quad Scale $d$ generalized eigenvectors proportionally such that all power constraints are satisfied;
\STATE \quad\quad Obtain $d$ achievable sum rates;
\STATE \textbf{End}
\STATE Return the generalized eigenvector which corresponds to the largest achievable sum rate.
\end{algorithmic}
\end{algorithm}

\emph{Complexity Analysis}: For a given training channel, the computational complexity to solve the feasibility check problem is dominated by the generalized eigenvalue decomposition, with complexity $\mathcal{O}(d^3)$. Thus, the overall complexity of our proposed rate-profile learning-based algorithm is $\mathcal{O}(d^3M)$.

\section{Simulation Results}

In the simulations, we assume that all transmitters and receivers are uniformly distributed within a disk of radius 100 m. A minimum separation of 20 m is enforced between any two nodes. The active IRS is located at the center of the disk, 15 m above the ground. The channel vectors $\mathbf{g}_k$ and $\mathbf{h}_{k}^H$ follow independent Rician fading with a Rician factor of 5. The path loss exponents of $\mathbf{g}_k$ and $\mathbf{h}_{k}^H$ are both 2.2. The path loss at a reference distance of one meter is -30 dB. The noise powers at the active IRS and receivers are $\sigma^2_{\text{IRS}}=\sigma^2=-90$ dBm. The minimum achievable rate requirement for each transmitter-receiver pair is $R_{\min}=1.5$ bps/Hz. The transmission power of each transmitter is $P_{\nu}=20$ dBm. The per-element transmission power constraint at the IRS is $P_e=15$ dBm. The number of training channels is $M=60$. 

In Fig. 1, we compare the average achievable sum rate of our proposed rate-profile learning-based algorithm with that of the WMMSE algorithm, denoted as ``RPL" and ``WMMSE" in the legend, respectively, in the considered IA system. For benchmarking, the performance of the system without IA is also presented, denoted as ``w/o IA". From Fig. 1, it is found that our proposed IA system outperforms that without IA when $P_t$ is higher than 9 dBm. This is because of the minimum achievable rate requirement \eqref{q20b} for each transmitter-receiver pair. With IA, each transmitter-receiver pair is allocated a dedicated degree of freedom for transmission, whereas without IA, wireless resources tend to be exclusively allocated for the transmitter-receiver pairs with more favorable channel conditions. Fig. 1 further confirms that our proposed algorithm outperforms the WMMSE algorithm. The latter is known to converge to a local optimum, and its performance is highly sensitive to the initial point. With an unfavorable initialization, it may be trapped in a suboptimal region. By contrast, our proposed learning-based algorithm, despite being designed for a suboptimal solution, exhibits robustness against local optimum entrapment.

\begin{figure}
\centering
\includegraphics[width=3.6in]{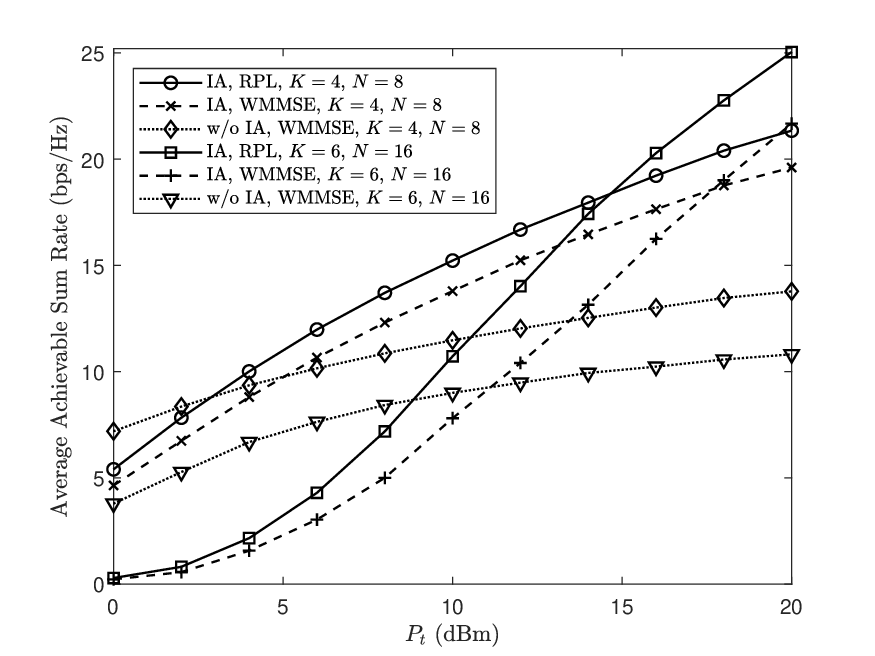}
\caption{Average achievable sum rate versus $P_t$; comparison of our proposed rate-profile learning-based algorithm with IA, the WMMSE algorithm with IA, and the system without IA.}
\end{figure}

In Fig. 2, we present the program execution time of our proposed rate-profile learning-based algorithm with IA, the WMMSE algorithm with IA, and the system without IA, where $K=6$ and $N=16$. All simulations were implemented in Python 3.12.10 using CVXPY 1.9.2, and executed on an AMD Ryzen 5 7500F CPU. From Fig. 2, it is observed that our proposed algorithm achieves a speedup of about 260 times over the WMMSE algorithm.

\begin{figure}
\centering
\includegraphics[width=3.6in]{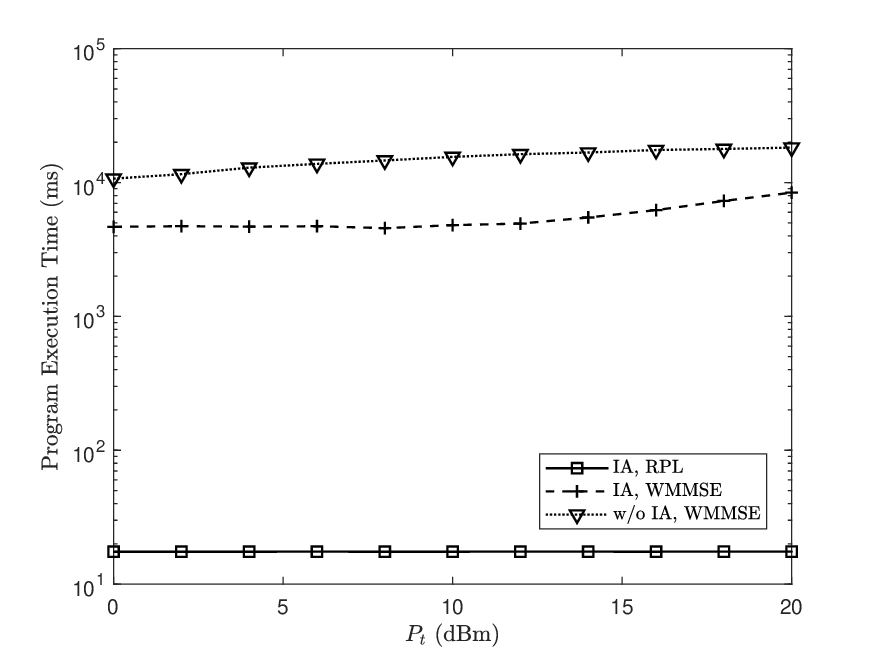}
\caption{Program execution time versus $P_t$; comparison of our proposed rate-profile learning-based algorithm with IA, the WMMSE algorithm with IA, and the system without IA, where $K=6$ and $N=16$.}
\end{figure}

\section{Conclusion}

In this letter, we have proposed a real IA scheme for an active IRS-aided system and developed a rate-profile learning-based algorithm as an alternative to the conventional WMMSE approach. Simulation results show that our proposed IA system outperforms that without IA when the IRS transmission power is sufficiently high. Furthermore, our learning-based algorithm achieves a higher sum rate than the WMMSE algorithm, while consuming significantly less execution time.

\appendices
\section{Proof of Proposition 1}

To satisfy $\mathbf{s}^T\mathbf{P}\mathbf{s}=P_t$, from \eqref{q15}, we have
\begin{align}\label{aq1}
\sum_{n=1}^N q_n \left(\left(\mathbf{z}_n^T\mathbf{s}\right)^2+\left(\mathbf{z}_{N+n}^T\mathbf{s}\right)^2\right)=P_t
\end{align}
where
\begin{equation}
\mathbf{Z}=\left(\begin{array}{c}
\mathbf{z}_1^T \\
\vdots \\
\mathbf{z}_{2N}^T 
\end{array}
\right)\in\mathbb{R}^{2N\times d}.
\end{equation}
To satisfy $\mathbf{s}^T\mathbf{Q}_n\mathbf{s}= P_e$, from \eqref{q13}, we obtain
\begin{align}\label{aq4}
q_n \left(\left(\mathbf{z}_n^T\mathbf{s}\right)^2+\left(\mathbf{z}_{N+n}^T\mathbf{s}\right)^2\right)=P_e.
\end{align}

In the following, we prove Proposition 1 by contradiction. Suppose the total power constraint is active, and at least one of the per-element transmission power constraints in \eqref{bq5b} is also active, which means that \eqref{aq1} is satisfied and \eqref{aq4} holds for a specific $n\in\mathcal{N}$. Subtracting both sides of \eqref{aq4} from those of \eqref{aq1}, we obtain
\begin{align}\label{aq10}
\sum_{n'=1,n'\neq n}^N q_{n'} \left(\left(\mathbf{z}_{n'}^T\mathbf{s}\right)^2+\left(\mathbf{z}_{N+n'}^T\mathbf{s}\right)^2\right)=P_t-P_e.
\end{align}
Given $\mathbf{s}$, the terms $\left(\mathbf{z}_{n'}^T\mathbf{s}\right)^2+\left(\mathbf{z}_{N+n'}^T\mathbf{s}\right)^2$ are constants. Since $q_{n'}$ includes the random channels $g_{n',k}$, the left-hand side of \eqref{aq10} is a random variable with a continuous probability density function. The probability that \eqref{aq10} holds tends to zero. This contradicts the assumption that the total power constraint is active, and at least one of the per-element transmission power constraints in \eqref{bq5b} is also active.

Suppose at least two of the per-element transmission power constraints in \eqref{bq5b} are active, i.e., \eqref{aq4} is satisfied for $n=n_1$ and $n=n_2$, where 
$n_1,n_2\in\mathcal{N}$ and $n_1\neq n_2$. Therefore, we have
\begin{align}\label{aq20}
\frac{\left(\mathbf{z}_{n_1}^T\mathbf{s}\right)^2+\left(\mathbf{z}_{N+{n_1}}^T\mathbf{s}\right)^2}{\left(\mathbf{z}_{n_2}^T\mathbf{s}\right)^2+\left(\mathbf{z}_{N+{n_2}}^T\mathbf{s}\right)^2}
=\frac{q_{n_2}}{q_{n_1}}.
\end{align}
Given $\mathbf{s}$, the left-hand side of \eqref{aq20} is a constant. The right-hand side of \eqref{aq20}, which includes  
the random channels $g_{n,k}$, is a random variable with a continuous probability density function. The probability that \eqref{aq20} holds tends to zero. This contradicts the assumption that at least two of the per-element transmission power constraints in \eqref{bq5b} are active.

\end{document}